\documentclass[a4paper,useAMS,usenatbib]{mn2e}

\usepackage{ifthen,psfig,graphicx}
\usepackage{epstopdf,aas_macros}
\usepackage{soul,color,amsmath}
\usepackage{rotating,longtable,caption,subcaption}
\usepackage{dblfloatfix}
\usepackage{placeins}
\def\mathnew{\mathsurround=0pt}
\def\simov#1#2{\lower 2.5pt\vbox{\baselineskip0pt \lineskip-.5pt
\ialign{$\mathnew#1\hfil##\hfil$\crcr#2\crcr\sim\crcr}}}
\def\simless{\mathrel{\mathpalette\simov <}}
\def\simgreat{\mathrel{\mathpalette\simov >}}
\newcommand{\MeV}{Me\kern-0.11em V}
\newcommand{\keV}{ke\kern-0.11em V}

\newcommand{\cha}{{\it Chandra\/}}

\newcommand{\ecmss}{erg~cm$^{-2}$ s$^{-1}$}
\newcommand{\ecmssa}{erg~cm$^{-2}$ s$^{-1}$ \AA$^{-1}$}
\newcommand{\es}{erg~s$^{-1}$}
\newcommand{\omegam}{$\Omega_{m,0}$\/}
\newcommand{\omegal}{$\Omega_{\lambda,0}$\/}
\newcommand{\Lsun}{\ensuremath{{\rm L}_{\odot}}}

\newcommand{\Mbh}{\ensuremath{M_{\bullet}}}
\newcommand{\Msun}{\ensuremath{{\rm M}_{\odot}}}

\makeatletter
\newcommand\ion[2]{\mbox{#1$\;${\small\expandafter\@slowromancap\romannumeral #2@\relax}}}
\newcommand\iont[2]{{#1$\;${\small\expandafter\@slowromancap\romannumeral #2@\relax}}}
\makeatother
\makeatletter
\newcommand{\raisemath}[1]{\mathpalette{\raisem@th{#1}}}
\newcommand{\raisem@th}[3]{\raisebox{#1}{$#2#3$}}
\makeatother

\title[Deep Spectroscopy of the $M_V\sim-14.8$ Host Galaxy of a Tidal Disruption Flare in A1795]{Deep Spectroscopy of the $M_V\sim-14.8$ Host Galaxy of a Tidal Disruption Flare in A1795\thanks{Based on observations obtained at the Gemini Observatory, which is operated by the Association of Universities for Research in Astronomy, Inc., under a cooperative agreement with the NSF on behalf of the Gemini partnership: the National Science Foundation (United States), the National Research Council (Canada), CONICYT (Chile), the Australian Research Council (Australia), MinistŽrio da Ci\^{e}ncia, Tecnologia e Inova\c{c}\~{a}o (Brazil) and Ministerio de Ciencia, Tecnolog\'{i}a e Innovaci\'{o}n Productiva (Argentina). }
}
\author[W. P. Maksym et al.]{W. P. Maksym$^{1}$\thanks{E-mail: wpmaksym@bama.ua.edu},
 M. P. Ulmer$^{2,3}$,
 K. C. Roth$^{4}$,
 J. A. Irwin$^{1}$,
R. Dupke$^{1,5,6,7}$,
\newauthor 
 L. C. Ho$^{8,9}$,
  W. C. Keel$^{1}$
  and C. Adami$^{10}$\\
$^{1}$University of Alabama, Department of Physics and Astronomy, Tuscaloosa AL 35487, USA\\
$^{2}$Northwestern University, Department of Physics and Astronomy, Evanston IL 60208, USA\\
$^{3}$Northwestern University, CIERA, Evanston IL 60208, USA\\
$^{4}$Gemini Observatory, Hilo, HI 96720, USA \\
$^{5}$Observatorio Nacional, Rua Gal. Jose Cristino, 20921-400 Rio de Janeiro, Brazil \\
$^{6}$Department of Astronomy, University of Michigan, 930 Dennison Building, Ann Arbor, MI 48109-1090, USA\\
$^{7}$Eureka ScientiÞc Inc., Oakland, CA 94602-3017, USA\\
$^{8}$Kavli Institute for Astronomy and Astrophysics, Peking University, Beijing 100871, China\\
$^{9}$The Observatories of the Carnegie Institution for Science, 813 Santa Barbara Street, Pasadena, CA 91101, USA\\
$^{10}$Laboratoire d'Astrophysique de Marseille, Marseille 13388, France
}
\begin{document}

\date{In original form 2013 December 10}

\pagerange{\pageref{firstpage}--\pageref{lastpage}} \pubyear{2013}

\maketitle

\label{firstpage}

\begin{abstract}

A likely tidal disruption of a star by the intermediate-mass black hole (IMBH) of a dwarf galaxy was recently identified in association with Abell 1795.  Without deep spectroscopy for this very faint object, however, the possibility of a more massive background galaxy or even a disk-instability flare from a weak AGN could not be dismissed.  We have now obtained 8 hours of Gemini spectroscopy which unambiguously demonstrate that the host galaxy is indeed an extremely low-mass ($M_\star\sim3\times10^{8}\;\Msun$) galaxy in Abell 1795, comparable to the least-massive galaxies determined to host IMBHs via other studies.  We find that the spectrum is consistent with the X-ray flare being due to a tidal disruption event rather than an AGN flare.  We also set improved limits on the black hole mass ($\rm{log}[\Mbh/\Msun]\sim5.3-5.7$) and infer a 15-year X-ray variability of a factor of $\simgreat10^4$.  The confirmation of this galaxy-black hole system provides a glimpse into a population of galaxies that is otherwise difficult to study, due to the galaxies' low masses and intrinsic faintness, but which may be important contributors to the tidal disruption rate.  

%We also discuss the origins of broad ($\sim1000\;\rm{km\;s}^{-1}$), weak ($L_{\lambda5007}\sim8.7\times10^{37}\;$\es) [\ion{O}{3}] emission and other possible weak high-ionization lines.

\end{abstract}


\section{Introduction}

%Black hole distribution

Massive black holes (MBHs; mass \Mbh$\simgreat10^{5}\;$\Msun) are thought to occur in most if not all galaxies and are thought to be an integral component of galactic formation and evolution.  MBHs become progressively more difficult to observe at lower masses, however, due to their diminishing influence on the surrounding stellar population and weaker accretion capability.  In addition, the broad emission-line region of active galactic nuclei (AGNs) may be strongly regulated for $\Mbh\simless10^6\;\Msun$ \citep{Chakravorty14}.  Such intermediate-mass black holes (IMBHs; $10^2-10^6\;\Msun$) are of great interest for the purpose of studying the physical nature of these relationships, and for devising models which can successfully predict the low-mass and high-mass tails of the black hole distribution \citep[e.g.][]{VN09}.  In particular, the presence of IMBHs in the dwarf galaxies of galaxy clusters is an issue of interest to models of galaxy formation and evolution, as these galaxies numerically dominate the overall galaxy population.

The disruption of a star by such an MBH may temporary produce a long-lived ($\sim$months--years) X-ray flare \citep{Rees88,Ulmer99} sufficiently luminous ($L_X\simgreat10^{42}\;$\es) to distinguish the emission from that of accreting stellar-mass black holes.  Such tidal disruption events (TDEs) in dwarf galaxies therefore provide an important complement to studies of low-mass AGNs \citep{Reines13} and weakly-accreting IMBHs in low-mass galaxies \citep{Miller12}, and hold implications for dynamical and evolutionary studies of galactic nuclei \citep[e.g.][]{Wang04}, including the rate of gravitational wave events from the disruption of white dwarfs by IMBHs \citep{Sesana08}.

\cite{Maksym13} recently identified a likely tidal disruption flare in a faint ($V=22.46$) galaxy associated with Abell 1795, WINGS J134849.88+263557.5 (hereafter WINGS~J1348) as part of a survey to find TDEs in archival X-ray observations of galaxy clusters \citep{Maksym10}.  \cite{Donato14} also later independently identified this flare and came to similar conclusions as to its origins.  These studies argued that WINGS~J1348 was likely a member of Abell 1795 with a projected distance of only $\sim50\;$kpc from the cluster core, and that an earlier {\it Extreme Ultraviolet Explorer} \citep[{\it EUVE};][]{EUVE} flare of indeterminate nature \citep[identified by][]{BBK99} had the same origins.  Such conclusions implied that the host galaxy was probably a dwarf galaxy ($M_V\sim-14.8$, $r\sim300\;$pc) hosting a possible IMBH ($\Mbh\sim3\times10^5\;\Msun$, from assumed Eddington accretion and from the $\Mbh-L_{bulge}$ relationship).  If so, it would be one of the least massive galaxies confirmed to host an IMBH, comparable to the most extreme examples from \cite{Reines13} and \cite{Miller12}.

The conclusions of \cite{Maksym13} and \cite{Donato14} rely directly on redshifts derived from template fitting of galaxy photometry and the likelihood of an associated galaxy being a true cluster member.  Both studies analyzed optical spectra which placed limits on any putative prior AGN which could produce an extreme X-ray flare.  But in the absence of such emission lines, neither spectrum was sufficiently deep to obtain a spectral redshift via absorption lines, given the faintness of the galaxy.  Seyfert 2 galaxies have been observed to commonly produce extreme long-term variability \citep{Saxton11AGN}, so WINGS~J1348 could in principle be a flaring Seyfert 2 at $z\simgreat0.4$ with weak emission lines \citep{Maksym13}.  Alternately, even if WINGS~J1348 has been quiescent until a recent TDE, it could be a background galaxy with significantly larger associated luminosities and \Mbh.  Finally, although \cite{Maksym13} and \cite{Donato14} both argue that a chance coincidence between the {\it EUVE} and \cha\ flare is inherently improbable, the field is sufficiently crowded that within the {\it EUVE} PSF there exist two brighter, unresolved SDSS objects ($V<21$; SDSS J134849.21+263550.5, SDSS J134850.01+263554.5) of uncertain nature which could conceivably be, e.g., flaring AGNs or Galactic objects.

In order to test these conclusions, we have obtained and analyzed deep (8 hours) Gemini multi-slit spectroscopy taken $\sim15$ years after the initial flare, which was necessary in order to confidently determine an absorption-line redshift.  In addition, given the excellent quality of the spectrum at redder wavelengths, we model the host galaxy via multi-component spectral template fitting.  We discuss the results of the spectral analysis to determine the classification of the host galaxy of the flare, as well as the two other bright optical objects within the {\it EUVE} PSF.  In addition, we discuss the implications for the flare itself.

%We also identify broad ($\sim1000\;\rm{km\;s}^{-1}$), weak ($L_{\lambda5007}\sim8.7\times10^{37}\;$\es) [\ion{O}{3}] emission, as well as several other possible weak high-ionization features, and speculate as to their origins and implications for the galaxy, the flare, and the MBH.  

Throughout this paper, we adopt concordant cosmological parameters of
$H_0=70\ $km$^{-1}$ sec$^{-1}$ Mpc$^{-1}$, \omegam=0.3 and \omegal=0.7,
and calculate distances using \cite{Wright06}. All coordinates are
J2000.

%Gas-accreting active galactic nuclei (AGNs) provide some of the best
%evidence for the existence of black holes, but despite the fact that
%AGN abundance appears greatest at $z \sim 2$ we have every reason to
%expect these black holes to play a role in their host galaxies beyond
%their transition into quiescence, although the population of MBH
%populations below $10^6\ \Msun$ is uncertain and it has been
%suggested that many less massive galaxies do not harbor an MBH at all
%\citep{FerrareseEtAl06}.

\section{Observations and Data}

\subsection{Observations}

We observed Abell 1795 from Gemini North with the {\it Gemini Multi-Object Spectrograph} \citep[GMOS;][]{GMOS} over 5 nights between 2013 April 8th and 2013 April 19th.  The queue science observations totaled 8 hours and were divided into sixteen 1800-second exposures.  We observed at least one Gemini Facility Calibration Unit (GCAL) flatfield per hour of exposure.  We used a slit mask with 58 slitlets of varying length, including WINGS~J1348, as well as nearby ($\sim 3.6$-arcsec and $\sim 11.1$-arcsec separation) objects SDSS~J134850.01+263554.5 and SDSS~J134849.21+263550.5 \citep[M13-A1795-S1 and M13-A1795-S2 from][]{Maksym13}, and 4 acquisition objects with {\it R}$\;\sim17-18$.  The other slitlets covered various other objects which will be discussed elsewhere (Maksym et al., {\it in prep.}).  The mask was designed from GMOS {\it r}-band pre-imaging taken in three 3-minute exposures on 2013 February 5.  Exposures were dithered in position and combined to compensate for chip gaps.
%% M13-A1795-S1 ID=1597
%% M13-A1795-S2 ID=1530
%% WINGS J1348 ID=1583

We used the B600 grating, which has an unbinned dispersion of $0.45\;\rm{\AA\;pixel}^{-1}$ and 2760\;\AA\ of simultaneous coverage.  We took 8 exposures per central wavelength at both 5200\;\AA\ and 5250\;\AA\ in order to compensate for the gaps between the GMOS chips.  This configuration afforded wavelength coverage at full exposure between 3824\;\AA\ and 6625\;\AA\ with rapidly declining sensitivity blueward of 4000\;\AA\footnote{http://www.gemini.edu/sciops/instruments/gmos/gratings/ gmos\_n\_B600\_G5303.txt}.  All slitlets covering science targets had 0.75-arcsec widths.  GMOS was centred on the coordinates of WINGS~J1348, and slitlets were positioned relative to GMOS pre-imaging as determined by {\small SEXTRACTOR} \citep{SExtractor96}, with WINGS~J1348 at $(\alpha,\delta)=(13^{h}48^{m}49^{s}.827$, $+26\degr35\arcmin57\arcsec.70)$, M13-A1795-S1 at $(\alpha,\delta)=(13^{h}48^{m}49^{s}.968$, $+26\degr35\arcmin54\arcsec.71)$, and M13-A1795-S2 $(\alpha,\delta)=(13^{h}48^{m}49^{s}.168$, $+26\degr35\arcmin50\arcsec.74)$.  Slit lengths and y-offsets were (7.00; 2.00), (3.70; -1.75) and (5.00; 0.00) respectively for optimal spacing.  GMOS was oriented 90 degrees east of north to minimize slit losses at high airmass, since GMOS is not equipped with an atmospheric dispersion corrector.  Readout was binned by two pixels in the spatial and spectral directions for an effective pixel size of 0.145 arcsec (0.91\;\AA\ per pixel), in order to improve the signal-to-noise ratio (SNR).  Conditions were better than 50th percentile for cloud cover and sky brightness, and 70th percentile for seeing (FWHM $< 0.6$ arcsec).  Average airmass was $\sim1.1$, but reached up to $\sim1.4$ for science observations.  

For absolute photometric calibration we used recently obtained GMOS spectra of standard star Feige 66 at 4200\;\AA, 5200\;\AA\ and 6200\;\AA, and for wavelength calibration we used CuAr arc lamps obtained the same night as the sub-exposures.  From CuAr arc exposures and sky lines of science exposures, we measure R$\;\sim1400$ for this grating configuration.  

\subsection{Data Reduction}

We processed our data according to the standard {\it Gemini} GMOS {\small IRAF} package and tools\footnote{http://www.gemini.edu/sciops/data-and-results/processing-software}.  We constructed bias files for both science and standard star observations using $\sim70$ bias observations obtained by GMOS for each configuration over a $\sim2$-week period encompassing the science observations.  We used the flat-field images to automatically edge-detect and cut the spectra.  Where the software failed to find sub-images (or ``strips") associated with individual spectra, we provided manual corrections.  We used the {\small IRAF} tool  {\small GSREDUCE} to remove bias, subtract the overscan values, and reject cosmic rays.  We calibrated the wavelength scale with {\small GSWAVELENGTH} and {\small GSTRANSFORM} by fitting fourth-order Chebyshev polynomials to the CuAr spectra.  We subtracted the sky emission by fitting first-order Chebyshev polynomials to the background with {\small GSSKYSUB}, and we then combined the data into two sets of spectra according to central wavelength by using {\small GEMCOMBINE}.  These spectra were flux-calibrated using {\small GSCALIBRATE} and sensitivity maps obtained from Feige 66 using {\small GSSTANDARD}.  Given the uncertainties in GMOS absolute photometric calibration for spectra, we folded our spectrum through the SDSS $r$-filter response in order to compare against our pre-imaging and Table 3 of \cite{Donato14}.  We find our standard star absolute calibration to be $\sim0.14$ magnitudes too bright, and we have recalibrated accordingly.

Using the {\small PROJECTION} region tool from {\small DS9}, we extracted spectra from a 0.85-arcsec strip centred on the brightest region.  We then combined the exposure-weighted spectra using {\small IDL}, interpolating the 5200\;\AA-centred spectrum to match the binning of the 5250\;\AA-centred spectrum.  Prior to combination, we eliminated instrumental features from chip gaps, as well as poorly subtracted emission from atmospheric [\ion{O}{1}] (5578.5\;\AA, 6301.7\;\AA) and Na ($\sim5890$\;\AA), substituting the mean continuum values of adjacent areas.  We identify one false emission line at $\sim4717$\;\AA\ which appears to be the result of cosmic rays intersecting over multiple exposures.
 We measure the SNR as $F_\lambda/\sigma_{F_\lambda}=(F_{5200}+F_{5250})/\sigma{\small(F_{5200}-F_{5250})}$ 
the standard deviation, where $F_{\lambda_c}$ is the flux for data with grating central wavelength $\lambda_c$, and $\sigma{\small(F_{5200}-F_{5250})}$ is the measured standard deviation of the flux difference between data taken with $\lambda_c=5200$ and $\lambda_c=5250$.  SNR is $\sim2$ near 4000\;\AA, increases approximately linearly to $\sim22$ at 6200\;\AA, and declines again to $\sim17$ near 6600\;\AA.  This estimate is comparable to our results from pre-observation simulations using the GMOS-N integration time calculator,\footnote{http://www.gemini.edu/sciops/instruments/integration-time-calculators/gmosn-itc} which assumed similar conditions.

The final, calibrated spectra, noise curves and residuals are plotted in Figures \ref{spectrum} and \ref{bluespec}.

%\begin{figure}
%\centering$
%\hspace{-0.2in}
%\begin{array}{l}
%\includegraphics[width=3.5in,angle=0]{fig1a.eps} \\
%\includegraphics[width=3.5in,angle=0]{fig1b.eps}
%\end{array}$
%\caption[]{Gemini spectrum of WINGS J1348, broken into below 5000\AA\ ({\it top}) and above 4900\AA\ ({\it bottom}).}
%\label{spectrum}
%\end{figure}

%\begin{figure*}
%\centering
%\begin{subfigure}{3.in}
%\includegraphics[width=5in]{fig1a.eps}
%\end{subfigure}

%\begin{subfigure}{3.in}
%\includegraphics[width=5in]{fig1b.eps}
%\end{subfigure}
%\caption[HST Image and Radial Profile of WINGS~J1348]{Left: WFPC2 image of WINGS~J1348, from F555W and F814W images combined to remove cosmic rays.  Right: Solid line, radial Profile of WINGS~J1348.  The y-axis is in units of magnitudes per square arcsecond for an annulus of mean radius corresponding to the x-axis.}
%\label{hstdata}
%\end{figure*}

\section{Analysis}

\subsection{Pre-imaging}

From the guide stars, we use the radial profile fitting function in {\small ATV} \citep{ATV} to infer a pre-imaging seeing of $\simless0.67\;$arcsec.  For WINGS~J1348, we measure $r=22.16$ (AB), and observed FWHM$\;=0.78\;$arcsec.  This $r$-value is consistent with values from Table 3 of \cite{Donato14}.  After correcting for seeing, the FWHM is comparable to the $\sim0.3$-arcsec optical FWHM of the dwarf galaxy as determined by \cite{Maksym13}.  The observed radial extent is $\sim1.0$ arcsec for a circular profile with 25.3 mag arcsec$^{-2}$ at $1\sigma$ above the background.  This result is comparable to the result determined by \cite{Donato14} from the same {\it HST} images analyzed by \cite{Maksym13}, and confirms the existence of fainter extended galactic structure.  M13-A1795-S1 is unresolved at $r=20.52\;$, and M13-A1795-S2 is resolved with observed FWHM$\;=0.84\;$arcsec and $r=20.24$.

\begin{figure*}
\includegraphics[width=7.25in]{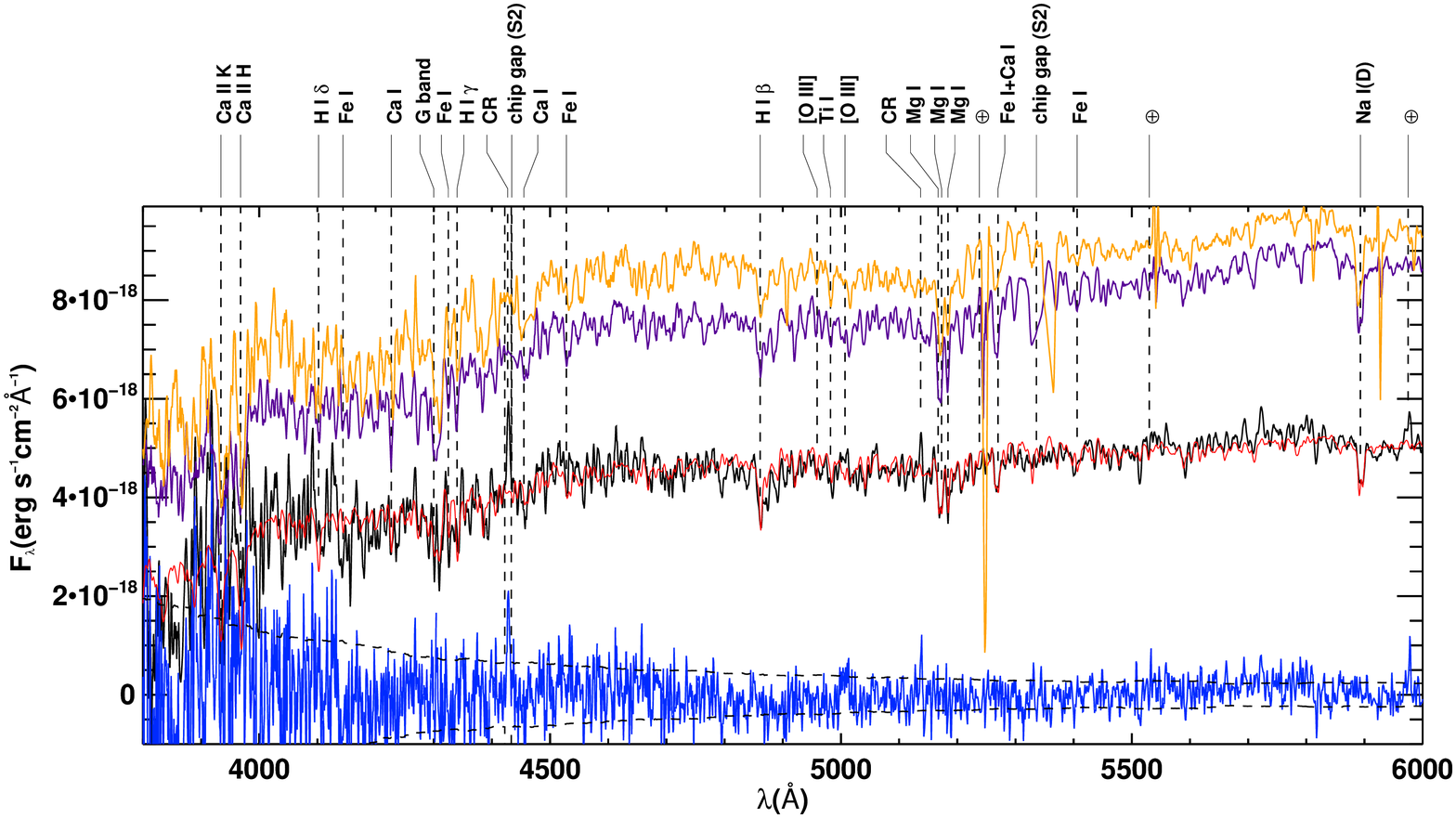}
\vspace{-0.5in}
\caption[]{Rest-wavelength Gemini spectrum of WINGS~J1348, with spectral features of interest indicated.  For clarity, the WINGS~J1348 data (black, solid) are smoothed with a boxcar average over five 0.9-\AA\ pixels.  The best-fit {\small PPXF} model (red) which includes \cite{CalzettiRed} internal reddening law are overlaid.  Near $\sim0\;$\ecmssa\ (bottom: blue, solid) is $F_\lambda(\rm{data})-F_\lambda(\rm{model})$.  Noise (black, horizontal dashed) represents local 1$\sigma$ uncertainty per $0.9-$\AA\ pixel, or alternately $3\sigma$ uncertainty for $\sim8-$\AA\ features.  A poorly subtracted cosmic ray (CR) near $4400\;$\AA\ has been is indicated as a noise spike.  The atmospheric feature near $6000\;$\AA\ is [\ion{O}{1}] $\lambda6363\;$\AA.  The model is good match for 
\ion{Mg}{1}~b ($\lambda5173\;$\AA),  \ion{Na}{1}~D ($\lambda5893\;$\AA), the G band ($\lambda4300\;$\AA) and numerous stellar absorption lines.  Rest-frame spectra from nearby early-type galaxies M13-A1795-S1 (top, orange; $z=0.0615$) and M13-A1794-S2 (second from top, purple; $z=0.0632$) have been rescaled to match WINGS~J1348, smoothed and overplotted with an offset for comparison.  Note that these comparison spectra use $\lambda_c=5250\;$\AA\ data only, and thus have not had deep absorption-like features from GMOS chip gaps removed.
A colour version of this figure is available in the online edition.
}
\label{spectrum}
\end{figure*}

\subsection{Absorption Line Identification and Continuum Modelling}\label{idmodel}

We consider the possibility that WINGS~J1348 is a background AGN, and initially identify no obvious emission lines in the spectra by which we may determine a redshift.  For a hypothetical background AGN, the 1-sigma limit on a FWHM~$=1000\;\rm{km\;s}^{-1}$ emission line is well-described as $F\simless(1.2\times10^{-17}\times[\lambda/(4000\;\rm{\AA})]^{-7}+1.6\times10^{-18})\;$\ecmss.  The spectrum appears devoid of strong emission features.

In \cite{Maksym13}, we argued that WINGS~J1348 was likely a member of A1795.  If so, Ca H+K $\lambda\lambda$3935,4214\;\AA\ would be close to 4200\;\AA, where the SNR remains relatively low.  In the case of a dwarf galaxy, the absorption lines might also be intrinsically weak.  Having previously used correlation analysis to investigate whether our 30-minute spectrum from the Magellan Echellete \citep{Mage} was consistent with this hypothesis, we do so again with the new, deeper {\it Gemini} data.  This time, we use {\small ZFIND} from the {\small SPEC2D} spectral analysis suite for {\small IDL}\footnote{http://www.astro.princeton.edu/\textasciitilde schlegel/code.html}, which cross-correlates the galaxy spectrum with that of a smooth eigenspectrum from the {\small SPEC2D} template library.

{\small SPEC2D} finds a family of minima with $1.09\simless\chi^2/\mu\simless1.12$ at $0.050\simless z\simless 0.067$, which is comparable to A1795 ($z=0.062$).  It finds a second family of minima with $1.07\simless\chi^2/\mu\simless1.10$ at $0.90\simless z\simless 0.94$, and a third family with $1.08\simless\chi^2/\mu\simless1.13$ at $1.08\simless z\simless 1.11$.

Since the $0.050\simless z\simless 0.067$ result is consistent with the redshift of A1795, we examine the spectra directly to see if typical small-scale absorption features (not present in the smooth eigenspectrum) correspond to the data near $z\sim0.062$.  Ca H+K are difficult to identify unambiguously due to the presence of multiple nearby features of comparable significance.  But H$\beta$ $\lambda4861\;$\AA\ and the \ion{Mg}{1}~b $\lambda5173\;$\AA\ triplet are prominent with continuum SNR~$\sim12$.  The \ion{Na}{1}~D $\lambda5893\;$\AA\ doublet  occurs at continuum SNR~$\sim22$.  These features require an adjusted $z=0.065$.  When fixed at this redshift, other less prominent features become obvious, including Ca H+K, several Fe features, and even the shape of the CN $\lambda3883\;$\AA\ band  \citep[see, e.g.][]{TB90}.  The high quality of the continuum becomes evident when compared to a significantly brighter early-type galaxy (SDSS J134842.56+263700.6, $M_g=17.55$) from the same observation set, whose continuum closely corresponds to that of WINGS~J1348 at scales up to hundreds of Angstroms.  

We confirm this analysis via multi-component fitting of single stellar population models.  We have modelled the spectra and measured the absorption line kinematics using {\small STARLIGHT} \citep{starlight} and Penalized Pixel Fitting ({\small PPXF}) software \citep{PPXF}.  For {\small STARLIGHT}, we use \cite{BC03} synthetic spectra spanning a grid of 15 ages up to 13 Gyr, and 3 metallicities where Z=\{0.004, 0.02, 0.05\}, and an expanded grid of 23 ages up to 13 Gyr and Z=\{0.0001, 0.0004, 0.004, 0.008, 0.02, 0.05\}.  Initial {\small STARLIGHT} models overpredict the continuum by $\sim10\%$ at $\lambda\simless4600\;$\AA\ and $\lambda\simgreat6000\;$\AA\ (rest), and require reddening $A_V\simless-0.9$ as per \cite{CalzettiRed}.  When we apply a polynomial correction to re-calibrate the Gemini data with SDSS J134842.56+263700.6 spectra from SDSS DR10, we find that this issue disappears.  We therefore use the re-calibrated data for our modelling. 

\begin{figure}
\includegraphics[width=3.3in]{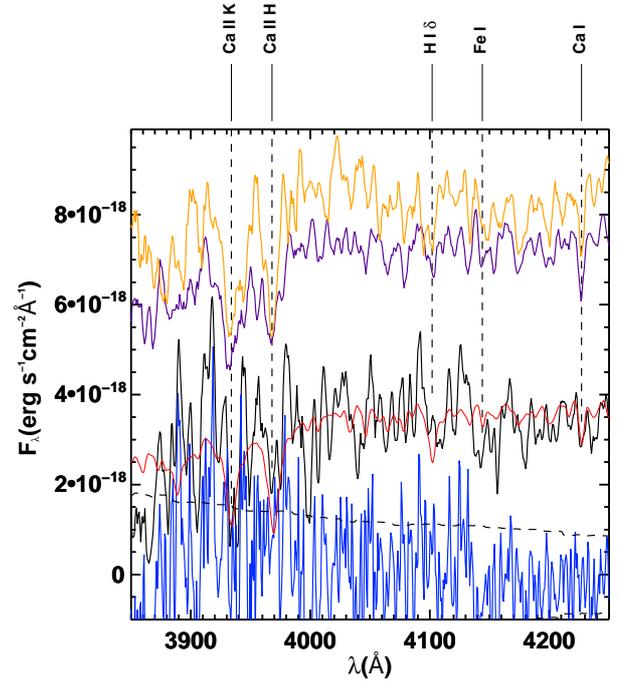}
\vspace{-0.5in}
\caption[]{As Figure \ref{spectrum}, but covering only 3800-4000\;\AA\ in order to more clearly display the Ca H+K absorption features.  Offsets of M13-A1795-S1 and M13-A1795-S2 have been displaced vertically for clarity.
}
\label{bluespec}
\end{figure}

For {\small PPXF}, we use \cite{miles} synthetic spectra spanning a grid of 48 ages up to 14 Gyr, with [M/H]=\{+0.22, 0.00, -0.40, -0.71, -1.31, -1.71\}.  These template spectra are convolved with the quadratic difference between the Gemini instrumental resolution and \cite{miles} template resolution.  We explore  {\small PPXF} models which include either \cite{CalzettiRed} reddening or which include polynomial corrections to the continuum fit.  We find the intrinsic reddening is negligible.  The best-fit {\small PPXF} model with \cite{CalzettiRed} reddening is overlaid in Figures \ref{spectrum} and \ref{bluespec}.

\begin{figure}
\includegraphics[width=3.15in]{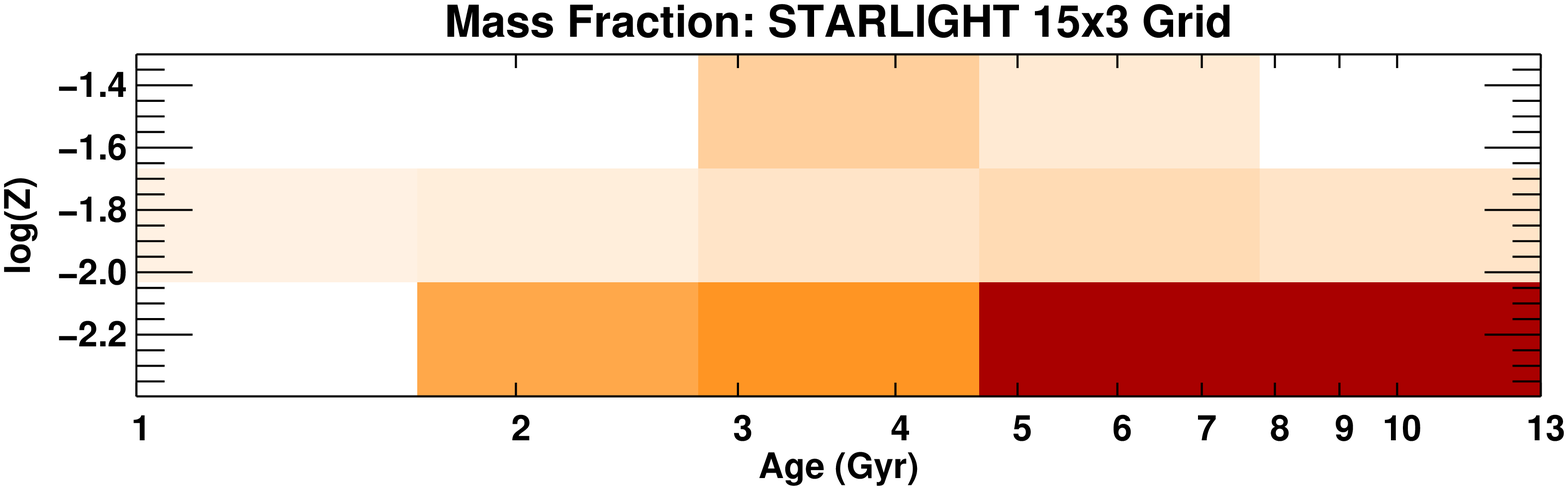}\\
\includegraphics[width=3.15in]{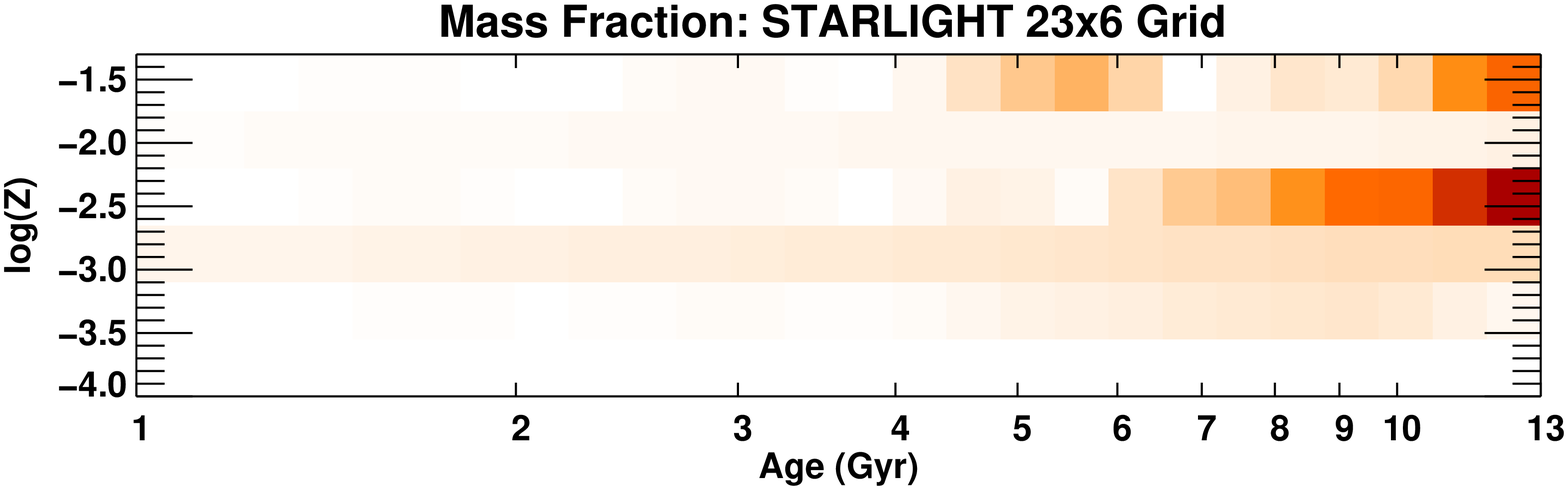}\\
\includegraphics[width=3.15in]{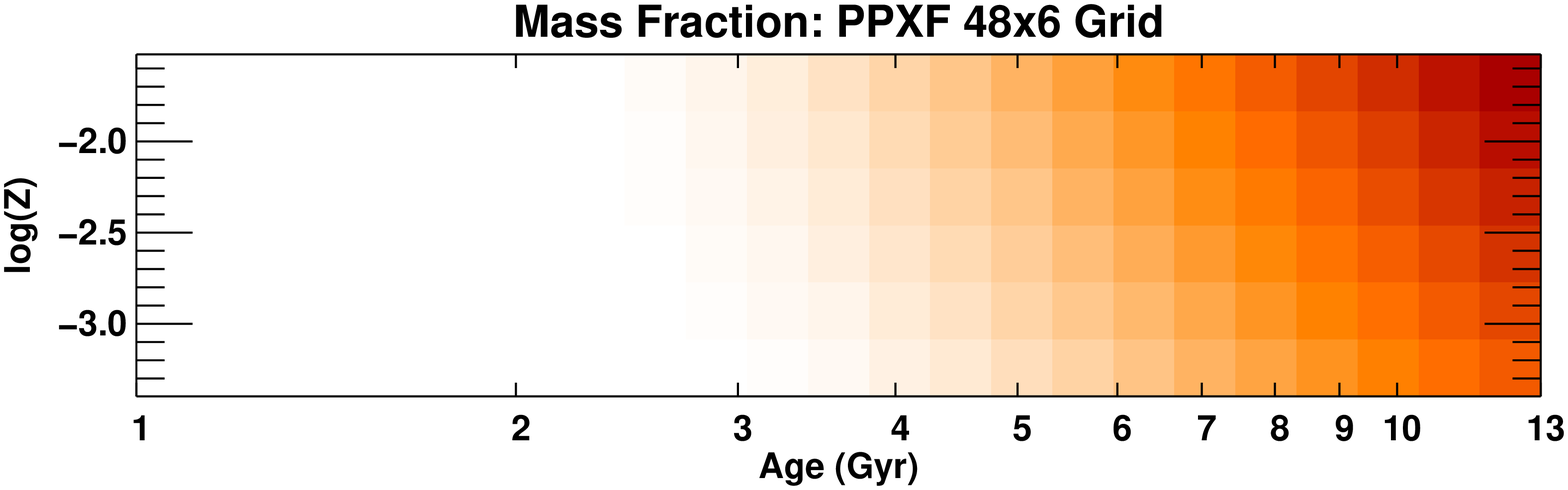}\\
\caption[]{Relative mass fractions of different stellar populations with respect to metallicity and age, as per the best-fit models from {\small STARLIGHT} and {\small PPXF} template synthesis codes.  Dominant populations are indicated by shading, with darker shading indicating a larger fraction of the total stellar mass in the best-fit model.  Populations with age $<1\;$Gyr are excluded due to relative lack of significance.
%In {\small PPXF}, [M/H]=\{0.22, 0.00, -0.40, -0.71, -1.31, 1.71\} corresponds to Z=\{0.03, 0.019, 0.008, 0.004, 0.001, 0.0004\}, or $-1.52\le \rm{log(Z)} \le-2.40$.
}
\label{popfit}
\end{figure}

The results of these fits, illustrated in Figure \ref{popfit}, are consistent with $\simgreat90\%$ of the stellar mass composed of an old ($\simgreat 5\;$Gyr), passive, low-metallicity (Z$\;\simgreat 0.004$) population, and possibly also containing a smaller, younger ($\sim1\;$ Gyr) component with higher metallicity (Z$\;\sim0.05$).   {\small STARLIGHT} also predicts a stellar mass ($M_\star$) of only $\sim(2.3-3.0)\times10^8\;$\Msun\ for the current stellar population.  All models find best-fit radial velocity estimates which are within one spectral dispersion element of $z=0.065$, such that $z=0.06514\pm0.00020$.  Furthermore, all models require velocity dispersions below the instrumental dispersion, $\sim50\;\rm{km\;s}^{-1}$ at $\sim5500\;$\AA.  This is consistent with our estimates that the stellar absorption lines are unresolved.  Since determinations of velocity dispersions below an instrumental dispersion are unreliable, we take the instrumental dispersion to be an upper limit.

%WR1: 
%WR2: 5762-5812

\subsection{Emission Line Measurement}\label{linemeasure}

Despite the high quality of the continuum, we do not identify any unambiguous emission features relative to either comparison galaxy spectra in the same field or the {\small PPXF} best-fit model.  For the purposes of determining the presence of any weak ongoing nuclear or star formation activity, we set upper limits on relevant diagnostic lines covered by our spectrum: [\ion{O}{2}] $\lambda3727\;$\AA, H$\beta$ $\lambda4861\;$\AA\ and [\ion{O}{3}] $\lambda5007\;$\AA.

In order to determine the strength of any faint narrow emission lines which are undetected above the continuum, we must assume a line width.  The simplest scenario supposes that the velocity of gas in the narrow line region is comparable to the stellar velocity dispersion \citep[$\rm{FWHM}_{gas}\sim0.8\;\rm{FWHM}_{\star}$, as per][]{Ho09}.  We therefore assume an upper limit of $\sigma_{gas}\simless50\;\rm{km\;s}^{-1}$, according to the instrumental dispersion and \S\ref{idmodel}.  

We subtract the best fit {\small PPXF} model with reddening from the measured spectrum and fit the difference to a gaussian with vertical offset.  The inferred $3\sigma$ upper limits are thus: $F_{\rm{[OIII]}}<2.3\times10^{-18}\;$\ecmss\ or $L_{\rm{[OIII]}}<2.1\times10^{37}\;$\es, $F_{\rm{H}\beta}<9.7\times10^{-19}\;$\ecmss\ or $L_{\rm{H}\beta}<9.7\times10^{36}\;$\es, and $F_{\rm{[OII]}}<5.4\times10^{-18}\;$\ecmss\ or $L_{\rm{[OII]}}<5.0\times10^{37}\;$\es.  We expect H$\alpha$ $\lambda6563\;$\AA\ to typically be stronger where Balmer emission is present ($F({\rm{H}\alpha})/F({\rm{H}\beta})\sim3.5$ in AGNs), but $\lambda6563\;$\AA\ lies outside our spectral range.  

Note that any emission lines from a typical AGN would likely be even narrower (and hence weaker) than this.  If, based the virial theorem, we assume $M_\star\sim5\sigma_\star^2 r_e/G$, then $\sigma_\star\simless38\;\rm{km\;s}^{-1}$.  Alternately, according to the \cite{FJ76} relation as extended by \cite{Jiang11b} to low-mass galaxies, we derive $\sigma_\star\sim27\;\rm{km\;s}^{-1}$ \citep[assuming $V-I=1.34\;$mag and $B-I=1.8$, as per][]{GHB08}.  There is significant uncertainty to these assumptions, however.  In addition to an intrinsic scatter of 0.07 dex to the results of \cite{Jiang11b}, their relation is poorly constrained for galaxies so intrinsically faint, with only one galaxy at $M_I>-16.50$ ($M_V>-15.16$).

These assumptions may be complicated by a possible broad excess observed near [\ion{O}{3}] $\lambda5007\;$\AA.  At only $\sim1\sigma\;\rm{pixel}^{-1}$, the SNR of this excess is strongly model-dependent ($\sim3\sigma-5\sigma$) and from inspection requires FWHM$\sim1000\;\rm{km\;s}^{-1}$ ($\sim17\;$\AA\ rest), which is strongly inconsistent with expectations for WINGS~J1348 (as per this section).  Due to its low significance, we defer further discussion of this putative broad feature to Appendix \ref{o3discuss}.  The primary impact of such a broad [\ion{O}{3}] $\lambda5007\;$\AA\ feature upon our main findings would be to suggest a more conservative upper limit for [\ion{O}{3}] $\lambda5007\;$\AA, such that $F_{\rm{[OIII]}}<1.5\times10^{-17}\;$\ecmss\ or $L_{\rm{[OIII]}}\sim1.3\times10^{38}\;$\es.  But in any event, such emission is unlikely to be caused by illumination of the narrow-line region given the constraints discussed in this section.

\subsection{Spectroscopy of M13-A1795-S1 and M13-A1795-S2}

M13-A1795-S1 and M13-A1795-S objects are $\sim1.5$ magnitudes brighter than WINGS~J1348 in $B$, $V$ \citep{Varela09} and $R$ according to the Gemini photometry.  As a result we easily identify their prominent Ca H+K absorption lines, as well as other features typical of early-type galaxies with high confidence.  These other objects are passive early-type galaxies common to galaxy clusters, they are members of A1795 ($z=0.0615$ and $z=0.0632$ respectively), and they have no evidence of strong line emission typical of AGNs.  We have rescaled their rest-frame spectra and plotted them in Figures \ref{spectrum} and \ref{bluespec} for comparison against WINGS~J1348.

%% Starlight stellar mass: 3.6*10^7 Msun (BN) 3.9*10^7 Msun (BS)

\section{Discussion}

From these observations, we can immediately infer several important results with respect to \cite{Maksym13}.  First and foremost, we unambiguously determine $z=0.06514\pm0.00020$ (heliocentric) for WINGS~J1348.  The host galaxy for the \cha\ 1999-2005 flare described in those papers is therefore clearly a member of A1795 ($z\sim0.062$).  \cite{Maksym13} had previously supported cluster membership via cross-correlation analysis of a 30-minute Magellen Echellete spectrum and association of the galaxy with the cluster's colour-magnitude ridge line.  WINGS~J1348 has a moderately high projected relative velocity of $\sim900\;\rm{km\;s}^{-1}$.  This is modestly high relative to the cluster dispersion \citep[$\sim658\;\rm{km\;s}^{-1}$,][]{Cava09}.  Give a projected distance of $\sim50\;$kpc and a first-order assumption of linear orbits for the cluster galaxies, this relative velocity is thus dynamically consistent with a galaxy near the cluster core.  This redshift determination also confirms the distance scales and luminosity distances determined photometrically by \cite{Maksym13}, as well as the inferred properties of the galaxy and associated flare: WINGS~J1348 is indeed a passive $M_V=-14.8$ dwarf galaxy \citep[see also][]{Donato14}.  
Via ATV \citep{ATV} profile fitting of the 1999 {\it HST} F555W observation and the analysis of \cite{Donato14}, we infer a half-light radius of $r_e\sim160\;$pc.

In addition, the unremarkable early-type spectra of M13-A1795-S1 and M13-A1795-S2 strongly support the hypothesis of \cite{Maksym13} that the \cha\ and {\it EUVE} flare originate from the same event, given there are no other comparably bright objects in the {\it EUVE} PSF core \citep[as per][]{Maksym13}.  Neither object is an AGN or galactic object which could conceivably produce such a bright flare except via another (and comparably rare) X-ray transient.

From both sets of best-fit {\small STARLIGHT} models to WINGS~J1348, we infer a stellar mass of $M_\star\sim2.5\times10^{8}\;$\Msun, which is somewhat larger than (but comparable to) $\sim8.6\times10^{7}\;$\Msun\ as derived according to \cite{dJB03} and the $B-R$ colour \citep[from, e.g., Table 3 in][]{Donato14}.  By comparison, Henize 2-10 has $M_\star\sim3.9\times10^{9}\;\Msun$ \citep{Reines11} and the LMC has $M_\star\sim2.7\times10^{9}\;\Msun$ \citep{vanderMarel06}.  $M_\star$ for WINGS~J1348 is more directly comparable to the least-massive galaxies in \cite{Reines13} or to M60-UCD1 ($M_\star\sim2\times10^8\;\Msun$; \citealt{Strader13}), though WINGS~J1348 is $>2$ orders of magnitude less dense.  At $L_V\sim10^{8}\;\Lsun$ and $\Sigma\sim10^{2.7}\;\Lsun\;\rm{pc}^{-2}$, WINGS~J1348 appears near the gap between ultra-compact dwarf galaxies and more luminous dwarf galaxies \citep[Figure 3. in][]{Strader13}, and given its $r\simgreat600\;$pc stellar envelope could represent a transition object, as per \cite{Brodie11}.  At a basic level, template fitting reveals a very old stellar population of uncertain or mixed metallicity, as suggested by \cite{Maksym13}.  The star formation rate inferred from our limit to H$\beta$ emission as per \cite{Moustakas06} is negligible at $\simless4.4\times10^{-4}\;\rm{\Msun yr}^{-1}$ and strongly reduces the likelihood of a core-collapse supernova explanation for the flare.  With so little ongoing star formation, WINGS~J1348 is unlikely to be a dwarf irregular and therefore may be a dwarf spheroidal.  But without better structural and kinematical information, any more precise label than ``dwarf galaxy" cannot be applied with certainty.

An assumed range of $10^8\;\Msun\simless\;M_\star\simless3\times10^8\;\Msun$ allows us to derive an expected \Mbh, assuming the stellar mass resides entirely in a spheroid.  In this case, the $M_{bulge}-\Mbh$ relationship of \cite{KH13} implies $\rm{log}(\Mbh/\Msun)\sim5.1-5.7$, which is consistent with \cite{Maksym13} and \cite{Donato14}.  This range is also consistent with an upper limit on \Mbh\ based on the best-fit kinematics of the absorption line spectrum, which imply an unresolved, lower-mass black hole.  Given $\sigma_\star\simless50\;\rm{km\;s}^{-1}$, we expect $\Mbh\simless7\times10^{5}\;\Msun$ \citep{KH13}.  With greater confidence in the {\it EUVE} flare's TDE origin, we also infer $\Mbh\simgreat2\times10^{5}\;\Msun$, assuming the TDE is Eddington-limited.

% Spec B 22.36 R 22.14 B-R 0.213
% WINGS B 23.28

We strongly constrain any persistent nuclear activity from WINGS~J1348, and improve upon previous limits from \cite{Maksym13} and \cite{Donato14}.  In particular, we set a limit on narrow H$\beta$ a factor of $\sim180$ lower than \cite{Maksym13}.  Even with our most conservative upper limits on weak [\ion{O}{3}] emission, we also infer that any nuclear activity must be temporary or very weak.  By comparison, \cite{Panessa06} and \citet[Fig. 3]{Ho12} imply $L_X(0.5-2.0\;\rm{keV})\sim1.5\times10^{39}\;$\es, or a factor of $\sim2\times10^4$ below the 1998 {\it EUVE} flux, and $\sim20$ below the 2010 \cha\ upper limit in \cite{Maksym13}.  We derive comparable results from \cite{SL12}, $L_X\sim2.6\times10^{39}\;$\es and $L_{bol}\sim6.9\times10^{39}\;$\es.  The inferred low level AGN emission is reduced even further if we assume narrower [\ion{O}{3}] widths as per \cite{Ho09} and \S\ref{linemeasure}, such that $L_X$ and $L_{bol}$ are reduced by factors of $\sim6$ and $13$ respectively.  Our spectrum does not cover the useful BPT \citep{BPT81} lines redward of 6300\;\AA, but we infer that the H$\beta$ is consistent with the \cite{Maksym13} nondetection of H$\alpha$ assuming $F({\rm{H}\alpha})/F({\rm{H}\beta})\sim3.5$.  

\section{Conclusions}

Via new deep Gemini spectroscopy we have obtained a redshift for the faint ($V\sim22.5$) host of the X-ray flare identified in archival \cha\ data by \cite{Maksym13} and later \cite{Donato14}, and have conducted a detailed analysis of its spectral properties.  We strongly confirm the major assertions of \cite{Maksym13} and \cite{Donato14}, namely that:

\begin{itemize}
\item Our derived redshift agrees with the initial redshift estimate from \cite{Maksym13}.  The flare therefore occurred in a dwarf galaxy in Abell 1795, with only $M_V\sim-14.8$.  The host galaxy, WINGS~J1348, has a half-light radius of $r_e\sim160\;$pc and an extended stellar envelope of $\simgreat600\;$pc.
\item All luminosities from \cite{Maksym13} and \cite{Donato14} based upon an assumed redshift of $z\sim0.062$ are also supported.
\item The luminous ($\simgreat3\times10^{43}\;$\es) {\it EUVE} flare of March 1998 is likely to have the same origin as the \cha\ flare, given the two brighter objects within $\sim15$ arcsec of WINGS~J1348 are both non-active early-type members of A1795. 
\item A tidal flare explanation is strongly favored, given limits on AGN emission which suggest no more than weak ($\simless10^{39}\;$\es) ongoing accretion, as well as a passive early-type galaxy spectrum with marginal star formation and an old ($\simgreat5\;$Gyr) stellar population.
\item The host galaxy is therefore one of the smallest galaxies confirmed to host an MBH, and is likely to host an intermediate-mass black hole.  Limits from $M_\star$ ($\rm{log}[\Mbh/\Msun]\sim5.1-5.7$) are consistent with those from Eddington and stellar dispersion limits ($\rm{log}[\Mbh/\Msun]\sim5.3-5.8$).
\item The TDE rate therefore may be favorably affected by the presence of IMBHs in cluster dwarf galaxies, as per \cite{Wang04}.  A larger dataset, however, is necessary to strongly constrain the rate from such objects.
\end{itemize}

In addition, we find:

\begin{itemize}
\item The implied total decline from the peak X-ray flux is a factor of $\simgreat10^4$ over a period of $\sim15\;$years.
\item WINGS~J1348 is moving at $\sim900\;\rm{km\;s}^{-1}$ relative to A1795, dynamically consistent with a linear orbit in the cluster core at a projected distance of $\sim50\;$kpc.  
\item The MBH is unlikely to have been ejected from an MBH-MBH merger given its low stellar dispersion ($\simless50\;\rm{km\;s}^{-1}$) and \citep[as per][]{Maksym13} large stellar mass indicative of a dwarf galaxy.  

\end{itemize}

These results support the presence of IMBHs in normally inactive dwarf galaxies, and complement efforts to understand the MBH population of low-mass galaxies such as via observations of AGNs \citep{Reines13} and weakly accreting black holes \citep{Miller12}.

Potential opportunities for future investigation include deep spectroscopy of WINGS~J1348 at $\simgreat6300\;$\AA\ to better constrain emission from [\ion{O}{1}], H$\alpha$,  [\ion{N}{2}], and  [\ion{S}{2}] diagnostic lines.  Also, deep observations at higher spectral resolution could better determine $\sigma_\star$ and hence directly compare the IMBH of this extremely low-mass galaxy with the $\Mbh-\sigma$ relation.  Such measurements would be of interest for comparison against measured environmental effects in the \Mbh\ relations, such as by \cite{McGee13}. 

\section*{Acknowledgments}

We thank the anonymous referee for helpful comments which improved the quality of the paper.

Data were obtained under Gemini program GN-2013A-Q-19.  We thank the Gemini staff for their help in planning and executing these observations.  WPM, JAI and WCK acknowledge support from the University of Alabama.  WPM acknowledges support from a University of Alabama Research Stimulus Program grant.  MU acknowledges partial support from Northwestern University.  LCH acknowledges support from the Kavli Foundation, Peking University, the Chinese Academy of Sciences, and the Carnegie Institution for Science.  WPM thanks Sean McGee for helpful conversation.  

This research has made use of the NASA/IPAC Infrared Science Archive, which is operated by the Jet Propulsion Laboratory, California Institute of Technology, under contract with the National Aeronautics and Space Administration.

Funding for SDSS-III has been provided by the Alfred P. Sloan Foundation, the Participating Institutions, the National Science Foundation, and the U.S. Department of Energy Office of Science. The SDSS-III web site is http://www.sdss3.org/.

SDSS-III is managed by the Astrophysical Research Consortium for the Participating Institutions of the SDSS-III Collaboration including the University of Arizona, the Brazilian Participation Group, Brookhaven National Laboratory, Carnegie Mellon University, University of Florida, the French Participation Group, the German Participation Group, Harvard University, the Instituto de Astrofisica de Canarias, the Michigan State/Notre Dame/JINA Participation Group, Johns Hopkins University, Lawrence Berkeley National Laboratory, Max Planck Institute for Astrophysics, Max Planck Institute for Extraterrestrial Physics, New Mexico State University, New York University, Ohio State University, Pennsylvania State University, University of Portsmouth, Princeton University, the Spanish Participation Group, University of Tokyo, University of Utah, Vanderbilt University, University of Virginia, University of Washington, and Yale University.

\bibliography{apj-jour,pete_tidal,biblio_mel_marc}
\bibliographystyle{mn2e}  

\FloatBarrier

\appendix
\section{A Possible Broad Excess Near \ion{O}{3} $\lambda5007\;$\AA}\label{o3discuss}

In \S\ref{linemeasure}, we mentioned a possible broad excess within $\sim9\;$\AA\ of \ion{O}{3} $\lambda5007\;$\AA.  We have noted that at only $\sim1\sigma\;\rm{pixel}^{-1}$ such a broad flux excess spans $\rm{FWHM}\sim1000\;\rm{km\;s}^{-1}$ if it is taken to be statistically significant, which is much greater than measured $\sigma_\star\simless50\;\rm{km\;s}^{-1}$ and therefore incompatible with standard assumptions regarding narrow line emission from AGNs \citep[see, e.g.,][]{Ho09}.

Emission from [\ion{O}{3}] is, however, predicted as long-term post-disruption feature by various TDE models \citep[e.g.][]{CE11,Clausen12,ELB95} and has been associated with TDE candidates \citep{Irwin10, CenkoNU}.   Such emission may be asymmetrical and comparably broad, owing to illumination of the tidal debris stream and ejecta, which continue to evolve at over long timescales and occupy spatial scales much smaller than the narrow line region of AGNs.  The detection of such features would therefore be of interest in comparison with such models.  We therefore describe it here for the sake of completeness.

\begin{figure}
\hspace{-0.2in}
\includegraphics[width=3.6in]{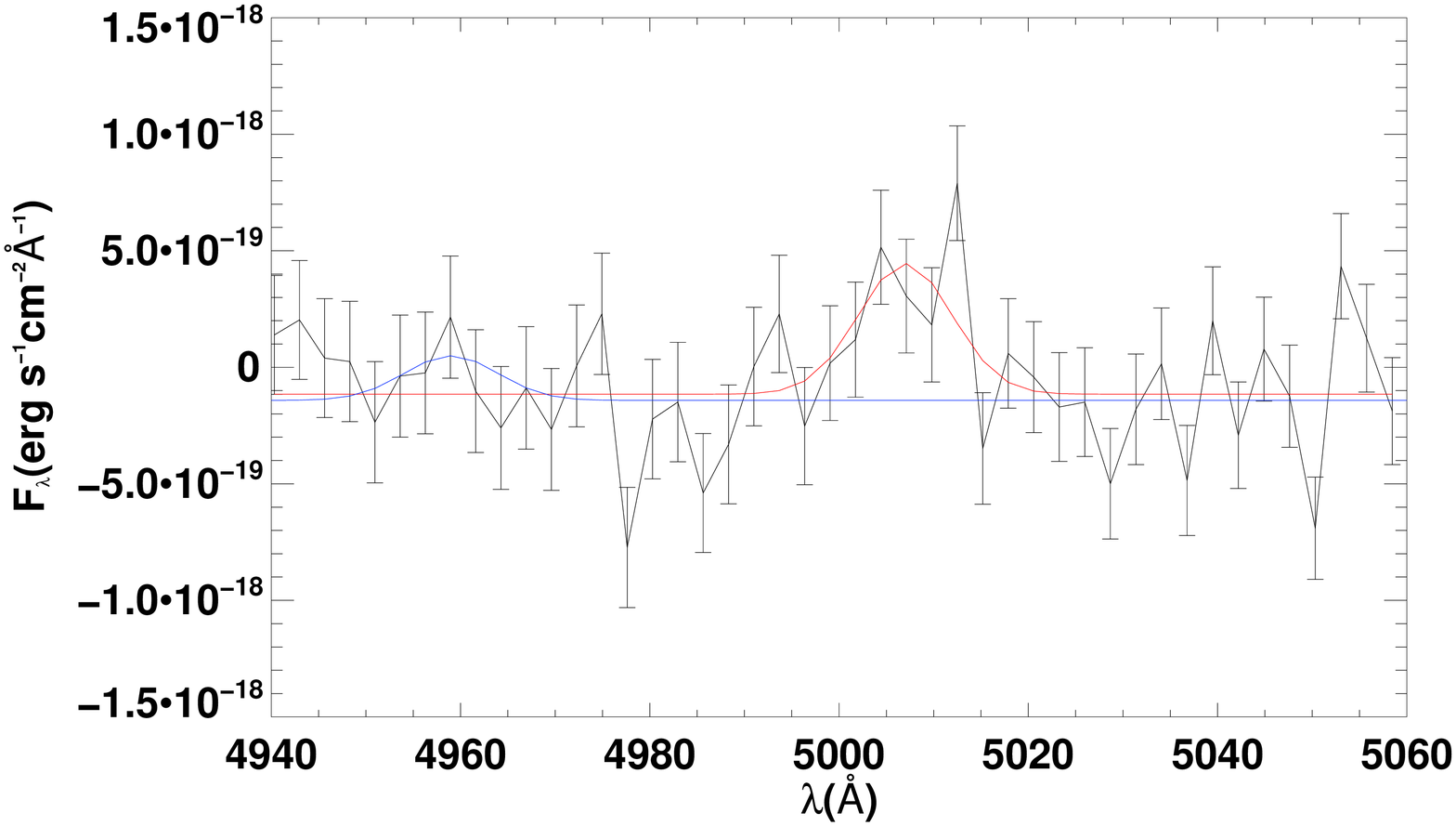}
\vspace{-0.1in}
\caption[]{$\lambda5007\;$\AA\ region (rest-frame), re-binned to 2.7\;\AA\ width for clarity.  Red: best-fit gaussian with vertical offset for a putative [\ion{O}{3}] line centred at $\lambda5007\;$\AA.  Blue: the same gaussian, centred at $\lambda4959\;$\AA\ and rescaled by 1/3. 
}
\label{o3fig}
\end{figure}

In Figure \ref{o3fig}, we display the region within 50\;\AA\ of $\lambda5007\;$\AA\ (rest-frame), re-binned to 2.7\;\AA\ width for clarity.  The statistical significance is insufficient to determine a width via gaussian fitting, so we estimate the FWHM from inspection of the unbinned data, with an uncertainty no less than $\sim200\;\rm{km\;s}^{-1}$.  Without accounting for systematic uncertainties in the continuum model, which are beyond the scope of this paper and may be significant, we find the putative detection to be $>3\sigma$ when fitting a simple gaussian to the unbinned data within 100\;\AA\ of $\lambda5007\;$\AA, whether using the best-fit {\small PPXF} model or by rescaling the early-type galaxy SDSS J134842.56+263700.6 from the same dataset to match the continuum.  When the gaussian model includes a vertical offset to match the {\small PPXF} continuum, $\sigma=4.8$.  We obtain comparable results when directly summing the flux excess.  

\FloatBarrier

Given possible systematic errors, the putative detection is therefore marginal and uncertain.  Also, any [\ion{O}{3}] $\lambda5007\;$\AA\ should be accompanied by [\ion{O}{3}] $\lambda4959\;$\AA\ with $\sim1/3$ of the $\lambda5007\;$\AA\ flux.  Such confirmation is not possible given the low SNR of [\ion{O}{3}] $\lambda5007\;$\AA\ (see Fig. \ref{o3fig}), however: any such $\lambda4959\;$\AA\ flux could be hidden by a $\simgreat1\sigma$ fluctuation.  There is no significant broad Balmer emission with the same FWHM, most notably for H$\beta$ $\lambda4861\;$\AA\ ($F_{H\beta}<5.5\times10^{-18}\;$\ecmss).  Since common emission lines are masked in the {\small PPXF} fit, such H$\beta$ is unlikely to be hidden by filling in the absorption line.

Considering the possibility that the excess could be spurious, we examined the sixteen sub-exposures individually and found no evidence for cosmic rays or instrumental artifacts.  We also compare against the co-added spectra from both $\lambda5200\;$\AA\ and $\lambda5200\;$\AA\ central wavelengths.  Each subset of eight exposures has an excess of $\sim1.3\times10^{-17}\;$\ecmss\ within 12\;\AA\ of 5007\;\AA, comparable to the best fit value for the full dataset.

A detailed treatment of the emission mechanics of the disruption is beyond the scope of this paper, but we note some basic constraints.  Assuming a cloud-and-stream model simplified from \cite{GMR14} to be stellar debris and a solar-type star disrupted at the tidal radius by an IMBH with \Mbh$\;\sim10^{5.4}\;$\Msun, after $\sim15$ years the marginally-bound material will have expanded to $\sim2\times10^{14}\;$cm, less than $r\sim3.4\times10^{15}\;$cm implied for Keplerian motion.  The area of the bound stream may be significant ($\sim2\times10^3\;\rm{AU}^2$), with a range of velocities $\simless9\times10^3\;\rm{km\;s}^{-1}$, and may include an outer sheath reaching the low critical density of [\ion{O}{3}] $n\sim6\times10^{5}\;\rm{cm}^{-3}$.  The data are insufficient, however, to demonstrate significant line-of-sight velocity which is likely from such a stream \citep{SQ09}.

Several interpretations of the putative [\ion{O}{3}] line are possible which do not necessitate gas dominated by the IMBH gravity or the dynamics of the debris cloud and ejecta stream.  For example, a radiatively-driven wind could occur at large ($\simless4.6$ pc) radii given an initial impulse from the early $\simgreat10^{43}\;$\es\ flare.  Such winds have been attributed to [\ion{O}{3}] emission from the Virgo elliptical galaxy NGC4472 globular cluster RZ 2109, possibly in conjunction with a stellar-mass black hole, and require relatively little gas mass \citep[$\simless10^{-3}\;\Msun$ for $L_{\lambda5007}\sim10^{37}\;$\es][]{Steele11}.  Such gas likely exceeds the escape velocity of the galaxy.  Note also that if the galaxy is moving through the cluster core, it is likely to be stripped of most of its gas, and what gas remains may be strongly influenced by the intracluster medium.  For comparison, $1000\;\rm{km\;s}^{-1}$ is comparable to the line-of-sight motion of WINGS~J1348 relative to A1795.  Also, the temperature-derived ICM sound speed is $\sim(5kT/3\mu m_p)^{1/2}\sim700-1300\;\rm{km\;s}^{-1}$ for $kT\sim2-6\;$keV plasma \cite[as measured by][]{Gu2012}, implying that the motion of WINGS~J1348 could be supersonic and might thus shock any gas still retained by the galaxy.

\bsp

\label{lastpage}

\end{document}